\documentstyle[psfig,pre,aps,twocolumn,float]{revtex}

\newcommand{\fr}[2]{\frac{\displaystyle{#1}}{\displaystyle{#2}}}
\newcommand{\Sum}[1]{\displaystyle \sum_{#1}}

\input{psfig}

\begin{document}
\draft
\title{\bf Field Driven Thermostated Systems : 
\\A Non-Linear Multi-Baker Map}
\author{T. Gilbert, C. D. Ferguson and J. R. Dorfman}
\address{Department of Physics and Institute for Physical Science and
Technology, \\ University of Maryland\\College Park MD, 20742, USA}
\date{\today}
\maketitle

\begin{abstract}
In this paper, we discuss a simple deterministic model for a field driven, thermostated
random walk that is constructed by a suitable generalization of a
multi-baker map.
The map is a usual multi-baker, but perturbed by a
thermostated external field that has many of the properties of the
fields used in systems with Gaussian thermostats. 
For small values of the driving field, the map is hyperbolic 
and has a unique SRB measure that we determine analytically to first order in 
the field parameter. We then compute the positive and 
negative Lyapunov exponents to second order and discuss their relation to 
the transport properties. For higher values of the 
parameter, this system becomes non-hyperbolic and posseses an attractive 
fixed point. 
\end{abstract}


\section{Introduction}

In the past several years a great deal of attention has been devoted
to computer and analytic studies of the chaotic properties of fluid
systems subjected to external fields and to Gaussian thermostats
which maintain a constant kinetic or total energy in the system, in
the presence of the field, \cite{hoo1,evmo}. The interest in this subject 
stems not only
from the method's value as a means of simulating non-equilibrium flows
and computing their properties, but also because there is a connection
between transport properties, non-equilibrium fluctuations, and the
underlying microscopically chaotic properties of the fluid. This
connection has been explored from computational \cite{poshoo,ecm1,ecm2,bec,mdr,md}
and analytic \cite{cels,rue1,btv,vtb,gd} points of view. The purpose of this 
paper is to describe a model
system in which the transport and dynamics of a thermostated system
can be studied in great detail, and in which one can explicitly
construct the SRB measure \cite{eckru} and describe such properties as the 
transition from hyperbolic to non-hyperbolic behavior, and related phenomena. 
These properties have been explored in previous work \cite{rue1,rue2},
but have not yet been
studied in great detail, due either to the complexity or to the
simplicity of the models treated up till now \cite{btv,vtb,gd}. The model 
discussed here
allows one to gain some insights into the general class of properties
of thermostated systems, while keeping the analytical and computational
difficulties to manageable proportions. It is one of the few 
cases known so far where one can check some of the general properties of 
thermostated systems on a specific model.

The model we consider is a variant of the multi-baker maps
studied by Gaspard and coworkers \cite{tasgas,gas1,gas2}, which are deterministic 
models for the diffusion of a particle on a one-dimensional lattice. The map 
considered here has, in addition, an external driving field which is 
constructed so as to model the effect of a thermostated electric field on 
charged particles in a two dimensional setting. 
We present the model and then calculate the chaotic
properties at small values of the external field. We obtain an
expression for the stationary state SRB measure to first order in the
applied field, and the positive and
negative Lyapunov exponents to second order in the applied field. This
allows us to verify the interesting relations between the rate of 
entropy production, the zero field
diffusion coefficient, the drift velocity, and the sum of the Lyapunov 
exponents \cite{evmo,rue1,vtb,gd}. We conclude with a brief discussion of
the 
transition to 
non-hyperbolic behavior as the field increases beyond a certain value, and 
discuss the connection of our model to other types of field-driven 
multi-baker maps \cite{btv,vtb,gd}.

\section{The non-linear multi-baker map}

\par We begin by considering a simple multibaker map that acts on the
$(x,y)$-coordinates of particles, and that models a random walk on a 
one-dimensional lattice of unit spacing. The map, defined on 
${\bf Z}\times[0, 1]^2$, replaces the $n, x, y$ coordinates of a particle
by $M_0(n, x, y)$ 
\begin{eqnarray}
M_0(n, x, y) &=& 
(n - 1, 2x, \fr{y}{2}),\hspace{1.5cm} 0 \leq x < 1/2,\nonumber\\
&=&(n + 1, 2x - 1, \fr{y + 1}{2}),\quad 1/2 \leq x < 1.
\label{scatmap}
\end{eqnarray}
Here $n$ represents the position of the random walker on the line and $x, y$ 
can be seen as bookkeeping variables keeping track of the
deterministic cause of the apparent random walk. The subscript, zero, on $M$ 
indicates 
that this is the map defined without an external field, which we introduce 
shortly. The map $M_0$ is time reversal symmetric. That is, there 
exists an involution operator, $T$, which acts on the $x,y$-variables, but
not on the box index $n$, and is given by $T(x, y) = (1 - y, 1 - x),$
with $T^2 ={\bf 1},$ where ${\bf 1}$ is the identity operator in 
${\bf R}^{2}$, and such that $T\circ M_0 \circ T (n, x, y) = M_0^{- 1}(n, x, y).$
If we consider periodic boundary conditions, the invariant measure is 
uniform and has Lyapunov exponents
\begin{equation}\label{0flyap}
\lambda^{(0)}_+ = - \lambda^{(0)}_- = \ln 2.
\end{equation}
\par Next we suppose that the particles
are also acted upon by a thermostated electric field, whose action we
now model. Our final map will then be a composition (to be explained below) of the field map (\ref{curtmap}) with the multi-baker 
map (\ref{scatmap}).
\par The modelling of the field map can be done by considering the action
of a thermostated electric field on a (continuously) moving
particle, where the thermostat maintains a constant kinetic energy for the particle.
The equation of motion of a particle in such a field is given by 
\cite{hoo1,evmo}
$$\fr{d \vec p}{d t} = q\vec E - q \left(\fr{\vec E\cdot\vec p}{p^2}\right)
\vec p.$$
If $\theta$ is the angle that the particle makes with respect to the 
direction of the electric field, then $\theta$ changes in time as 
$\dot{\theta}= - q E\sin\theta /p$,
with solution
$$\theta(t) = 2\arctan\left(\tan\left(\fr{\theta(0)}{2}\right)\exp\left(-\fr{qE}{p}t\right)\right).$$
A time discretized version of this equation is obtained by defining 
an angle at discrete times 
$\theta_n=\pi x_n$, and letting $x_n$ satisfy
$$x_{n + 1} = \fr{2}{\pi}\arctan\left(\tan\left(\fr{\pi x_n}{2}\right)e^{- \alpha}\right).$$
Here $x_n \in [0,1]$ and $\alpha = \fr{qE}{p}\tau,$ with $\tau$ the time
step, which for the time being we set equal to unity. Note that we restricted 
our attention to angles $\theta \in [0,\pi]$, 
taking advantage of the symmetry $\theta \leftrightarrow - \theta.$
\par We now introduce a one parameter family of maps of the unit
interval onto itself by
\begin{equation}\label{curtmap}
\varphi_\alpha(x) =  \fr{2}{\pi}\arctan\left(\tan\left(\fr{\pi x}{2}\right)e^{- \alpha}\right).
\end{equation}
We note the following property~: 
\begin{equation}\label{trfield}
\varphi_\alpha(x) = 1 - \varphi_{-\alpha}(1 - x),
\end{equation}
which implies the time reversal symmetry
\begin{equation}\label{trfield2} 
T\circ (\varphi_\alpha, {\bf 1})\circ T (x, y) = (x, \varphi_\alpha^{- 1}(y)),
\end{equation}
where $T$ is defined above, and the operator $(\varphi_\alpha, {\bf 1})$ acts on a point $(x,y)$, say to produce $(\varphi_{\alpha}(x),y)$, and we have denoted the identity operator acting on the $y$-coordinate by ${\bf 1}$. We also point out the fact that the map $\varphi_\alpha(x)$ has the property, under successive iterations, that
\begin{equation}
\varphi_{\alpha}(\varphi_{\alpha}(x))=\varphi_{2\alpha}(x),
\label{itmap}
\end{equation}
which shows that the field map is a good discretization of a continuous process in time.
\par We can now construct the following non-linear (or field driven) 
multi-baker map as a time reversal symmetric composition of the multi-baker 
map (\ref{scatmap}) with the field map (\ref{curtmap})~:
\begin{equation}
M_\alpha(n, x, y) = 
({\bf 1}, {\bf 1}, \varphi_\alpha)\circ M_0 \circ 
({\bf 1}, \varphi_\alpha, {\bf 1}) (n, x, y),
\label{!!}
\end{equation}
which takes the explicit form
\begin{eqnarray}
M_\alpha(n, x, y) &=& 
(n - 1, 2\varphi_\alpha(x), \varphi_\alpha(\fr{y}{2})),\nonumber\\
&&\hspace{3cm} 0 \leq x < \varphi_{-\alpha}(\fr{1}{2}),\nonumber\\
&=&(n + 1, 2\varphi_\alpha(x) - 1, \varphi_\alpha(\fr{y + 1}{2})),\nonumber\\
&&\hspace{3cm} \varphi_{-\alpha}(\fr{1}{2}) \leq x < 1.
\label{fieldbaker}
\end{eqnarray}
The leftmost identity operators in each of the field maps in (\ref{!!}) act on the cell index, $n$, and express the fact that the field maps do not change the value of the cell index. Only the baker map moves points from one cell to the next.
We refer to Fig. (5) at the end of this paper for an illustration of the projection 
of $M_\alpha$ along the $x$-interval.
The time reversal symmetry of this map $T\circ M_{\alpha} \circ T (x, y) = 
M_\alpha^{- 1}(x, y),$ with $T$ defined as above, follows straightforwardly
from (\ref{trfield}, \ref{trfield2}).
\section{Hyperbolic regime}
\par As long as $\alpha < \ln(2)$, $M_\alpha$ is expanding along the 
$x$-direction, i.e.
$$\fr{\partial M_{\alpha x}}{\partial x} > 1,$$
so that standard theorems guarantee the existence and uniqueness of an SRB 
measure \cite{beck}. In this section we solve for the invariant density and 
give an analytic expression of this invariant measure. 
\par We want to find the stationary eigenfunction, equivalently, the
invariant density, $\rho(n,x,y)$, of the Perron-Frobenius operator for a
system with periodic boundary conditions. This implies that $\rho$
does not depend upon $n$, but only on $x,y$  and 
satisfies the equation
\begin{eqnarray}
\lefteqn{\rho(x, y) =
\varphi'_{-\alpha}(\fr{x}{2})\varphi'_{- \alpha}(y)
\rho(\varphi_{- \alpha}(\fr{x}{2}), 2\varphi_{- \alpha}(y)),}\quad\nonumber\\
&&\hspace{4cm} 0 \leq y < \varphi_\alpha(\fr{1}{2}),\nonumber\\
&=& \varphi'_{-\alpha}(\fr{x + 1}{2})\varphi'_{- \alpha}(y)
\rho(\varphi_{- \alpha}(\fr{x + 1}{2}), 2\varphi_{- \alpha}(y) - 1),\nonumber\\
&&\hspace{4cm} \varphi_\alpha(\fr{1}{2}) \leq y < 1 
\label{PFgen}
\end{eqnarray}
where the prime denotes the derivation with respect to the argument.
We solve this equation by expanding the density in powers of the field 
parameter,
\begin{equation}\label{rho}
\rho(x, y) \simeq 1 + \alpha\rho^{(1)}(x, y) + \alpha^2\rho^{(2)}(x, y) 
+ o(\alpha^3).
\end{equation}
\par For small $\alpha$, the low field regime, we expand $\varphi_\alpha$ in powers of 
the field parameter :
\begin{equation}\label{lfcurtmap}
\varphi_\alpha(x) \simeq x - \fr{\alpha}{\pi}\sin(\pi x) + 
\fr{\alpha^2}{4\pi}\sin(2\pi x) + o(\alpha^3).
\end{equation}
The first order correction to the invariant density is found by decomposing
it in Fourier modes :
\begin{equation}\label{rho1}
\rho^{(1)}(x, y) = \sum_{k = 0}^\infty [a_k(y)\cos(2\pi k x) + 
b_k(y)\sin(2\pi k x)].
\end{equation}
Inserting the expansion for $\rho$, Eq. (\ref{rho}), and
Eq. (\ref{rho1}) in Eq. (\ref{PFgen}), we can find the $a_k$'s and
$b_k$'s. However, as a result of the phase space contraction, the density is a singular function of 
the $y$ coordinate so that we cannot represent the $a_k$'s and $b_k$'s in 
terms of standard functions \cite{tgd}. For our purposes, it is enough to 
perform a 
partial integration of the density along the $y$ direction so as to obtain 
continuous coefficients for the Fourier modes. We thus define
\begin{equation}
A_k(y) = \int_0^y a_k(y')dy';\,\,\,B_k(y) = \int_0^y b_k(y')dy'.\label{akAk}
\end{equation}
which, to lowest order in $\alpha,$ are found to satisfy the recursion 
relations
\begin{eqnarray}
A_0(y) &=& \fr{1}{2}A_{0}(2 y) + \fr{2 y}{\pi} + 
\fr{1}{\pi}\sin(\pi y) \nonumber\\ 
&&\mbox{} + \fr{1}{\pi}\Sum{k'\:{\rm odd}}\fr{B_{k'}(2 y)}{k'},\nonumber\\
&&\hspace{4cm} 0 \leq y < \fr{1}{2},\nonumber\\
&=& \fr{1}{2}A_{0}(2 y - 1) + \fr{2(1 - y)}{\pi} + \fr{1}{\pi}\sin(\pi y)
\nonumber\\ 
&& \mbox{} + \fr{1}{\pi}\Sum{k'\:{\rm odd}}\left(\fr{B_{k'}(1)}{k'} - 
\fr{B_{k'}(2 y - 1)}{k'}\right),\nonumber\\ 
&&\hspace{4cm}  \fr{1}{2} \leq y < 1,
\label{A0}
\end{eqnarray}
\begin{eqnarray}
A_k(y) &=& \fr{1}{2}A_{2k}(2 y)- \fr{y}{\pi(4 k^2 - 1/4)}\nonumber\\
&& \mbox{}  - \fr{2}{\pi}\Sum{k'\:{\rm odd}}\fr{k'}{4k^2 - k'^2}B_{k'}(2 y),\nonumber\\
&&\hspace{4cm} 0 \leq y < \fr{1}{2}\nonumber\\ 
&=& \fr{1}{2}(A_{2k}(1) + A_{2k}(2 y - 1)) 
- \fr{1 - y}{\pi(4 k^2 - 1/4)}\nonumber\\
&& \mbox{}  - \fr{2}{\pi} \Sum{k'\:{\rm odd}}\fr{k'}{4k^2 - k'^2}(B_{k'}(1) 
- B_{k'}(2 y - 1)),\nonumber\\ 
&&\hspace{4cm}  \fr{1}{2} \leq y < 1,\label{Ak}
\end{eqnarray}
\begin{eqnarray}
B_k(y) &=& \fr{1}{2}B_{2k}(2 y) + \fr{4 k y}{\pi(4 k^2 - 1/4)}\nonumber\\
&& \mbox{} + \fr{4 k}{\pi}\Sum{k'\:{\rm odd}}\fr{1}{4k^2 - k'^2}A_{k'}(2 y),\nonumber\\ 
&&\hspace{4cm}  0 \leq y < \fr{1}{2},\nonumber\\
&=& \fr{1}{2}(B_{2k}(1) + B_{2k}(2 y - 1)) 
+ \fr{4 k y}{\pi(4 k^2 - 1/4)}\nonumber\\
&& \mbox{} + \fr{4 k}{\pi}\Sum{k'\:{\rm odd}}\fr{1}{4k^2 - k'^2}
(A_{k'}(1) - A_{k'}(2 y - 1)), \nonumber\\
&&\hspace{4cm}  \fr{1}{2} \leq y < 1.
\label{Bk}
\end{eqnarray}
\par In particular, for $y = 1$, we find $A_k(1) = A_{2k}(1)$ and 
$B_k(1) = B_{2k}(1) + \fr{4k}{\pi(4k^2 - 1/4)},$ whose solutions are, 
respectively,
\begin{equation}
A_k(1) = 0\:;\quad B_k(1) = \fr{4}{\pi}\sum_{n = 0}^\infty\fr{2^n k}{2^{2(n + 1)}k^2 - 1/4}      \label{Ak(1)}
\end{equation}
Therefore the projection of $\rho^{(1)}(x,y)$ along the $x$-direction,
is a smooth function of $x$ given by
\begin{equation}\label{rho1x}
\rho^{(1)}(x) \equiv \int_0^1 \rho^{(1)}(x, y)dy = 
\sum_{k = 1}^\infty B_k(1)\sin(2\pi k x).
\end{equation}
To compute the coefficients $A_k(y)$ and $B_k(y)$ numerically, we need to
select a cut-off value $k_{max}$ of $k$ beyond which we set all the 
coefficients to be zero. In Figs. (1)-(3), we show $A_0$ and the first five 
$A_k$'s and $B_k$'s, respectively, computed by setting $k_{max} = 250.$  
\par Although we have found the invariant density only to first order,
we can now compute the corrections to the Lyapunov
exponents 
to second order in $\alpha$.
For $\lambda_+,$ we obtain
\begin{eqnarray}
\lambda_+ &=& \int_0^1 dx\int_0^1 dy \rho(x, y)\ln(2\varphi_\alpha'(x))
\nonumber\\
&=& \ln(2) - \alpha^2\left(\fr{1}{4} + \int_0^1 dx \rho^{(1)}(x)\cos(\pi x)
\right)\nonumber\\
&=& \ln(2) - \alpha^2\left(\fr{1}{4} + \fr{4}{\pi}\sum_{k = 1}^\infty
\fr{k B_k(1)}{4k^2 - 1}\right)\nonumber\\
&=& \lambda_+^{(0)} - \alpha^2(0.6661513 \pm 10^{- 7}).\label{plyap}
\end{eqnarray}
Here we used the normalization condition for $\rho(x,y)$ which
requires that  
\begin{equation}\label{consist}
\int_0^1 dx \int_0^1 dy \rho^{(2)}(x, y) = 0.
\end{equation}

To compute $\lambda_-,$ we need the full expression of $\rho^{(1)},$ 
Eqs. (\ref{rho1}-\ref{Bk}) :
\begin{eqnarray}
\lambda_- &=& \int_0^{\varphi_{-\alpha}(1/2)}dx\int_0^1 dy 
\rho(x, y)\ln(\fr{1}{2}\varphi_\alpha'(y/2)) \nonumber\\
&& \mbox{}  + \int_{\varphi_{-\alpha}(1/2)}^1 dx\int_0^1 dy 
\rho(x, y)\ln(\fr{1}{2}\varphi_\alpha'((y + 1)/2))
\nonumber\\
&=& \varphi_{-\alpha}(1/2)\left(-\ln(2) - \fr{2\alpha}{\pi} - \fr{\alpha^2}{4}
\right) \nonumber\\
&& \mbox{} - \alpha^2\int_0^{1/2}dx\int_0^1 dy\rho_1(x, y)\cos(\pi y/2)
\nonumber\\
&& \mbox{}  + (1 - \varphi_{-\alpha}(1/2))\left(-\ln(2) + \fr{2\alpha}{\pi} - \fr{\alpha^2}{4}
\right) \nonumber\\
&& \mbox{} + \alpha^2\int_{1/2}^1 dx\int_0^1 dy\rho_1(x, y)\sin(\pi y/2)
\nonumber\\
&=& \lambda_-^{(0)} - \alpha^2\left(\fr{1}{4} + \fr{4}{\pi^2} 
+ \fr{1}{\pi}\Sum{k\:{\rm odd}} \fr{B_k(1)}{k}\right.\nonumber\\
&& \mbox{} + \fr{\pi}{4}\int_0^1 dy A_0(y)(\sin(\pi y/2) + \cos(\pi y/2))  
\nonumber\\ 
&&\left.+ \fr{1}{2}\int_0^1dy \Sum{k\:{\rm odd}} \fr{B_k(y)}{k}
(\sin(\pi y/2) - \cos(\pi y/2)) \right).\nonumber\\
\label{nlyap}
\end{eqnarray}

\par It is not straightforward to compute these integrals numerically because 
of the irregularity of the functions (\ref{A0}-\ref{Bk}) and the number of 
different terms involved in their expressions. We can nevertheless
estimate (\ref{nlyap}) within some good accuracy. To this purpose, we proceed 
by a number of algebraic manipulations.
\par We first substitute for $B_k(y)$ the expression 
\begin{equation}\label{Bksubst}
B_k(y) = yB_k(1) + kf_k(y),
\end{equation}
where the functions $f_k(y)$ are found to satisfy the relations
\begin{equation}\label{fk}
f_k(y) = f_{2k}(2y) + \fr{4}{\pi}\sum_{k'\:{\rm odd}}
\fr{1}{4k^2 - k'^2}A_{k'}(2y),
\end{equation}
for $0 \leq y < 1/2,$ and $f_k(y)$ is odd with respect to $1/2,$ viz. 
$f_k(y) = - f_k(1 - y).$
\par In terms of these, we have
\begin{eqnarray}
\lefteqn{A_k(y) = \fr{1}{2}A_{2k}(2y) 
-\fr{2}{\pi}\sum_{k'\:{\rm odd}}\fr{k'^2}{4 k^2 - k'^2}f_{k'}(2y)}\quad&&
\nonumber\\
&&- \fr{y}{\pi}\left(\fr{4}{16k^2 - 1}
+ \sum_{k'\:{\rm odd}}\fr{4 k'}{4 k^2 - k'^2}B_{k'}(1)\right),
\label{newAk}
\end{eqnarray}
for $0 \leq y < 1/2,$ and $A_k(y)$ is even with respect to $1/2,$ viz. 
$A_k(y) = A_k(1 - y).$
\par We can now find an upper bound on the magnitude of $f_k(y).$ Indeed,
$A_k(y)$ is everywhere negative and is minimal at $y = 1/2$ (see Fig.(2)). 
Thus
\begin{equation}\label{boundAk}
|A_k| \leq \fr{1}{2\pi}\left(\fr{4}{16k^2 - 1}
+ \sum_{k'\:{\rm odd}}\fr{4 k'}{4 k^2 - k'^2}B_{k'}(1)\right).
\end{equation}
Now, $f_k(y)$ is negative between $0$ and $1/2$ and reaches its minimum
at $y = 1/4.$ Thus, from Eqs. ({\ref{fk}, \ref{boundAk}), 
\begin{eqnarray}
\lefteqn{|f_k| \leq f_k^{max} \equiv \fr{2}{\pi^2}\sum_{k'\:{\rm odd}}
\fr{1}{4k^2 - k'^2}}\hspace{1.5cm}\nonumber\\
&&\times\left(\fr{4}{16k^2 - 1}
+ \sum_{k'\:{\rm odd}}\fr{4 k'}{4 k^2 - k'^2}B_{k'}(1)\right).\label{boundfk}
\end{eqnarray}

\par Next, we rewrite $A_0(y)$, Eq. (\ref{A0}), in terms of its Fourier
modes~:
\begin{equation}\label{newA0}
A_0(y) = \fr{1}{\pi}\sum_{k = 0}^\infty G_k\cos(2\pi k y)
+ H_k\sin(2\pi k y).
\end{equation}
We find
\begin{eqnarray}
G_0 &=& 1 + \fr{4}{\pi} + \sum_{k'\:{\rm odd}}\fr{B_{k'}(1)}{k'}\nonumber\\
G_k &=& -\fr{4}{\pi^2 k^2}\left(1 + 
\sum_{k'\:{\rm odd}}\fr{B_{k'}(1)}{k'}\right) - \fr{4}{\pi(4 k^2 - 1)}
\nonumber\\
&& + 2\sum_{k'\:{\rm odd}}\int_0^1 dy f_{k'}(y)\cos(k\pi y),
\quad k \:{\rm odd},\nonumber\\
G_k &=& \fr{1}{2}G_{k/2} - \fr{4}{\pi(4 k^2 - 1)},
\quad k \:{\rm even},\nonumber\\
H_k &=& 0,\quad k \:{\rm odd},\nonumber\\
H_k &=& - \fr{4}{\pi k}\sum_{k'\:{\rm odd}}\fr{B_{k'}(1)}{k'}
+ 2\sum_{k'\:{\rm odd}}\int_0^1 dy f_{k'}(y)\sin(k\pi y),
\nonumber\\
&& \hspace{3cm}k \:{\rm even},\label{FourierA0}
\end{eqnarray}
\par Therefore and  with the help of Eqs.(\ref{boundAk}, \ref{boundfk}),
\begin{eqnarray}
\lefteqn{\int_0^1 dy \sum_{k\:{\rm odd}}\fr{B_k(y)}{k}(\sin(\pi y/2) 
- \cos(\pi y/2))}\quad&&\nonumber\\
&=&\fr{2}{\pi}\left(\fr{4}{\pi} - 1\right)\sum_{k\:{\rm odd}}\fr{B_k(1)}{k}
\nonumber\\
&&+ \sum_{k\:{\rm odd}}\int_0^1 dy f_k(y)(\sin(\pi y/2) - \cos(\pi y/2))
\nonumber\\
&=& \fr{2}{\pi}\left(\fr{4}{\pi} - 1\right)\sum_{k\:{\rm odd}}\fr{B_k(1)}{k}
+ O\left(\fr{4}{\pi}(\sqrt{2} - 1)f_k^{max}\right),\nonumber\\
\label{intsumBk}
\end{eqnarray}
and
\begin{eqnarray}
\lefteqn{\int_0^1 dy A_0(y)(\sin(\pi y/2) + \cos(\pi y/2))}\quad&&\nonumber\\
&=& \fr{4 G_0}{\pi^2} - \fr{4}{\pi^2}\sum_{k = 1}^\infty\fr{G_k}{16 k^2 - 1}
\nonumber\\
&=& \fr{4 G_0}{\pi^2} - \fr{4}{\pi^2}\sum_{k\:{\rm odd}}\left(
\sum_{n = 0}^\infty\fr{1}{2^n}\fr{G_k}{2^{2n + 4}k^2 - 1}\right.\nonumber\\
&&\hspace{2cm}\left. + \sum_{n = 1}^\infty\sum_{j = 0}^{n - 1}
\fr{1}{2^j}\fr{g_{2^{n - j}k}}{2^{2n + 4}k^2 - 1}\right)\nonumber\\
&=& \fr{4 G_0}{\pi^2} - \fr{4}{\pi^2}\sum_{k\:{\rm odd}}\left(
\sum_{n = 0}^\infty\fr{1}{2^n}\fr{G_k^{(0)}}{2^{2n + 4}k^2 - 1}\right.
\nonumber\\
&&\hspace{2cm}\left. + \sum_{n = 1}^\infty\sum_{j = 0}^{n - 1}
\fr{1}{2^j}\fr{g_{2^{n - j}k}}{2^{2n + 4}k^2 - 1}\right)\nonumber\\
&&+ O\left(\fr{4}{\pi^2}\sum_{k\:{\rm odd}}\sum_{n = 0}^\infty\fr{1}{2^{n - 1}}
\fr{1}{2^{2n + 4}k^2 - 1}\right.\nonumber\\
&&\hspace{3cm}\times\left.\fr{2}{\pi}\sum_{k'\:{\rm odd}}\fr{f_{k'}^{max}}{k'}
\right),
\label{intA0}
\end{eqnarray}
where we introduced the notations
\begin{equation}\label{Gk0}
G_k^{(0)} = -\fr{4}{\pi^2 k^2}\left(1 + 
\sum_{k'\:{\rm odd}}\fr{B_{k'}(1)}{k'}\right) - \fr{4}{\pi(4 k^2 - 1)}
\end{equation}
and
\begin{equation}\label{gk}
g_{2^{n - j}k} = - \fr{4}{\pi(2^{2(n - j + 1)}k^2 - 1)}.
\end{equation}
\par Grouping Eqs.(\ref{intsumBk}, \ref{intA0}) together with 
Eq.(\ref{nlyap}), we can give an estimate of the second order correction
to the negative lyapunov exponent by performing straightforward numerical
summations~: we find
\begin{equation}\label{nlyap_est}
\lambda_- = \lambda_-^{(0)} - \alpha^2(1.9937 \pm O(0.04)).
\end{equation}
\par In Fig. (4) 
we compare the values (\ref{plyap}) and (\ref{nlyap_est}) of the second order 
corrections to the Lyapunov exponents to numerically computed ones.

\par In the next section we take the macroscopic limit and relate the sum of 
the Lyapunov exponents to the drift velocity and the zero-field diffusion coefficient.

\section{Macroscopic limit}

\par In order to take the macroscopic limit of this diffusive process, we 
consider the function $W_{t}(n)$ which is defined to be the total probability 
of finding a particle in cell $n$ at time $t$, and is given by
\begin{equation}\label{Wn}
W_t(n) = \int_0^1dx\int_0^1dy \rho_t(n, x, y),
\end{equation}
where $\rho_t(n, x, y)$ is the time-dependent density of the extended version
of $M_\alpha$ and whose time evolution is derived by generalizing
Eq. (\ref{PFgen}) in a straightforward way. Once we have eliminated the 
internal variables, $(x,y)$, we no longer have a deterministic process, but 
instead have a random process which should, in a macroscopic limit, be 
described by a suitable Fokker-Planck equation with a drift term, representing
the effect of the external field. The macroscopic limit is taken by scaling 
the space and time parameters, and then taking an appropriate scaling limit. 
This procedure was described for models of this type by T\'el, Vollmer, and 
Breymann \cite{btv,vtb}, and here we simply outline the process. 
\par We have previously introduced the time step $\tau$, and we replace the 
times $t$ and $t+1$ in the Frobenius-Perron equation by $T\tau$ and 
$T\tau +\tau$, respectively, where $T >> 1$. Similarly, we scale the length 
of the elementary cells, $n$, by making them have a length $a$ on a side. We 
then replace $n$ and $n+1$ in the Frobenius-Perron equation by $Na$ and 
$Na +a$, respectively, where $N >> 1$. Also the $x$ and $y$ variables have to 
be scaled by the factor $a$, as well.
\par We can now proceed to the derivation of the Fokker-Planck equation from 
the Frobenius-Perron equation for $\rho_{t}(n,x,y)$. We obtain a Fokker-Planck 
equation for a field-driven random walk by considering the difference 
$W_{T\tau + \tau}(Na) - W_{T\tau}(Na)$.

Using Eqs. (\ref{fieldbaker},\ref{PFgen}), we find
\begin{eqnarray*}
W_{T\tau + \tau}(Na) &=& \int_0^{\varphi_{-\alpha}(1/2)}dx\rho_{T\tau}
(Na + a, x) \\
&& \mbox{} + \int_{\varphi_{-\alpha}(1/2)}^1 dx\rho_{T\tau}(Na - a, x),
\end{eqnarray*}
where $\rho_{T\tau}(Na, x) = a\int_0^1dy\rho_{T\tau}(Na, x, y).$ We have 
scaled $x$ and $y$ so that their values are in the interval 
$0 \leq x,y \leq 1$. 
Therefore,
\begin{eqnarray}
\lefteqn{W_{T\tau + \tau}(Na) - W_{\tau}(Na) =}\quad\nonumber\\ 
&& \int_0^{\varphi_{-\alpha}(1/2)}dx(\rho_{T\tau}(Na + a, x) - \rho_{T\tau}
(Na, x))
\nonumber\\
&& \mbox{} + \int_{\varphi_{-\alpha}(1/2)}^1 dx(\rho_{T\tau}(Na - a, x)
- \rho_{T\tau}(Na, x)).
\label{FPstep2}
\end{eqnarray}
\par Expanding $\rho_{T\tau}(Na \pm a, x)$ about $\rho_{T\tau}(Na, x)$ and 
$W_{T\tau+\tau}(Na)$ about  $W_{T\tau}(Na)$ respectively, and introducing the coordinate $X=Na$, and the time $t=T\tau$,
we get
\begin{eqnarray}
\lefteqn{\fr{\partial W_t(X)}{\partial t}=}\quad\nonumber\\
&& \fr{a}{\tau}\fr{\partial}{\partial X}
\left(\int_0^{\varphi_{-\alpha}(1/2)}dx \rho_t(X, x) - \right.
\nonumber\\
&&\left.\int_{\varphi_{-\alpha}(1/2)}^1 dx \rho_t(X, x)\right) 
+ a^{2}\fr{1}{2\tau}\fr{\partial^2 W_t(X)}{\partial X^2} +\cdots\nonumber\\
&=& - {\overline v}_t(X)\fr{\partial W_t(X)}{\partial X} + 
\fr{a^{2}}{2\tau}\fr{\partial^2 W_t(X)}{\partial X^2} +\cdots. \label{FPeq}
\end{eqnarray}
The {\em drift velocity} is given by,
\begin{eqnarray}
\lefteqn{{\overline v}_t(X) \equiv \fr{a}{\tau}\times}\nonumber\\
&& \fr{\fr{\partial}{\partial X}
\left(- \int_0^{\varphi_{-\alpha}(1/2)}dx\rho_t(X, x) + 
\int_{\varphi_{-\alpha}(1/2)}^1 dx \rho_t(X, x)\right)}
{\fr{\partial}{\partial X}\int_0^1 dx \rho_t(X, x)}.\nonumber\\
\label{dv}
\end{eqnarray}
To proceed further, we need to take the macroscopic limit where $a \rightarrow 
0, \tau \rightarrow 0, \fr{a^2}{2\tau} =  D$ and $\overline{v}$ is finite and 
nonzero. This implies that the electric field, $E$, becomes infinite as 
$a^{-2}$.

For long times, $t$, we are going to replace the 
drift velocity in the Fokker-Planck equation by its stationary state value. 
Suppose $t$ is large enough 
that we are nearing a steady state. Then in the limit of small $a$ and $\tau$, 
we may write the solution of the Frobenius-Perron equation $\rho_{t}(X,x)$ in 
(\ref{dv}) as $W_t(X)\rho(x)$ where $\rho(x)$ is determined by an equation 
easily obtained by integating the equation (\ref{PFgen}) over $y$ 
(\ref{rho1x}), and $W_t(X)$ is close to, but not quite, a constant. Then  the 
$X$ dependence drops out  in the expression for the drift velocity, and we 
find that the stationary state drift velocity is simply (assuming that the 
density is normalized to a unit cell)
\begin{equation}\label{driftv}
{\overline v} =\frac{a}{\tau} \left(
- \int_0^{\varphi_{-\alpha}(1/2)}dx\rho(x) + 
\int_{\varphi_{-\alpha}(1/2)}^1 dx \rho(x)\right).
\end{equation}

\par Eq. (\ref{FPeq}) is the Fokker-Planck equation corresponding to a
stochastic diffusive system with a drift. In the case of {\em periodic} boundary 
conditions that we consider here, we know from thermodynamics that the rate of
entropy production in the stationary state is due solely to the existence
of a current driven by the external field and for which the rate of 
entropy production is given by
\begin{equation}\label{epthermo}
\sigma = \fr{{\overline v}^2}{D},
\end{equation}
where the zero-field diffusion coefficient for this process is
$D = \fr{a^2}{2\tau}$. We mention that in the limit of zero field, and periodic boundary conditions, the entropy production in the stationary state vanishes, since in this limit, there is no steady state drift, and the distribution function $\rho(X,x)$ is constant both in $X$ and in $x$.  
\par With the stationary state density (\ref{rho1x}) computed in the previous
section, the drift velocity (\ref{driftv}) is found to be
\begin{eqnarray}
{\overline v} &=& -\fr{2a\alpha}{\pi\tau}\left(1 + \sum_{k\:{\rm odd}}\fr{B_k(1)}{k}
\right)\nonumber\\
&=& - \fr{a\alpha}{\tau}(1.15217813).\label{driftvalue}
\end{eqnarray}
Hence, the entropy production rate is, Eqs. (\ref{epthermo} - 
\ref{driftvalue}),
\begin{equation}\label{epvalue}
\sigma = 2.65502889\fr{\alpha^2}{\tau}.
\end{equation}
According to the usual arguments for thermostated systems, one expects
that the rate of phase space contraction given by the negative of the
sum of Lyapunov exponents should be equal to the macroscopic rate of
entropy production \cite{rue1,gd}. For our case, it is possible to verify 
this relation analytically, since we have been able to calculate all of 
the relevant quantities. The phase space contraction rate is given by
\begin{equation}\label{pscontrac}
- (\lambda_+ + \lambda_-) = (2.66 \pm O(0.04))\fr{\alpha^2}{\tau},
\end{equation}
and Eqs. (\ref{epvalue},\ref{pscontrac}) give consistant values. Thus,
our field driven random walk model has a well behaved macroscopic
limit, provides an example of the correspondence between the
macroscopic and microscopic relations for entropy production, and is
analytically tractable. 

\section{Conclusions and Discussion}

\par In this paper, we have shown that it is possible to construct a 
non-linear version of the multibaker map (i.e. it shares the topology and
time-reversal symmetry of the original multi-baker but is not piecewise
linear) that simulates the action of an external field on a diffusive 
process. The field curves the branches of the map and is responsible for the 
phase space contraction that induces a stationary state on an attractor (it
fills the whole phase space but its information dimension \cite{eckru} is 
fractional as a consequence of the difference between the second order 
corrections to the Lyapunov exponents).
\par One of the motivations of this work was to provide an analytically 
tractable map which shows some of the properties of the periodic Lorentz gas  
where the particle moves among the scatterers in a thermostated electic field. 
The structure of our non-linear baker map is sufficiently simple that we were 
able to compute analytically the stationary state SRB measure
using a perturbation expansion in the field parameter. This allowed us
to compute the positive and negative Lyapunov exponents, whose values we
showed are consistent with that of the drift velocity. We also were able to 
compute the irreversible entropy production and showed that it is indeed 
given, apart from some factors, by the sum of the Lyapunov exponents. 
\par We remark that at values of the field parameter larger than $\ln(2)$,
the map loses its hyperbolicity. This is illustrated in Fig. (5). In fact 
$(0,0)$ becomes an attractive
fixed point of the reduced map, which means that, on the lattice, all 
particles eventually move ballistically around the ring towards decreasing 
$n$'s. The case $\alpha = \ln(2)$ is of particular interest. Indeed, the
origin is an intermitent fixed point and, as a consequence, points can spend
arbitrary long times in its vicinity. This can be seen to give rise to 
anomalous diffusion \cite{sin,tasgas2}.
\par We regard the model given here as the simplest of a class 
of similar models which can be generated by varying an additional parameter
modelling a magnetic field. 
Elsewhere \cite{gd2} we will describe this class of models in much more 
detail, because they show a wide variety of features both in the 
hyperbolic and non-hyperbolic regions, including sequences of period adding 
bifurcations.

\par In a subsequent paper \cite{gd3} we will consider boundary conditions 
other than
periodic. It is in particular an important question to check whether, with 
flux boundary conditions, the chain sustains a stationary state with an 
almost linear gradient of density. As we showed in another paper \cite{gd}, 
some other models of field driven multi-baker maps fail to have 
this behavior in a large 
system limit and, as a consequence, do not have physically relevant 
thermodynamics (the entropy production rate depends on the choice of the
partition).
\par Another interesting perspective 
will be to calculate non linear corrections to the diffusion coefficient so 
we can go beyond the linear response theory. As shown by other authors 
\cite{gas2}, relevant tools for this study are the zeta functions and 
Pollicott-Ruelle 
resonances. Computing the first order corrections to the eigenvalue spectrum 
of the Perron-Frobenius operator will be an important step towards 
understanding non-linear diffusion for this model.

\begin{acknowledgments}
The authors wish to thank Brian Hunt, Celso Grebogi, Jurgen Vollmer, 
Tamas T\'el, Rainer Klages, Edward Ott,
Karol \.Zyczkowski and Mihir Arjunwadkar for helpful discussions, and 
Pierre Gaspard, Shuichi Tasaki, and Constantino Tsallis for pointing out the 
interesting intermittent behavior at the hyperbolic-nonhyperbolic transition. 
J. R. D. wishes to acknowledge support from the National Science Foundation 
under grant PHY -96 -00428.
\end{acknowledgments}

\newpage







\begin{figure}[htb]
\centerline{\psfig{figure=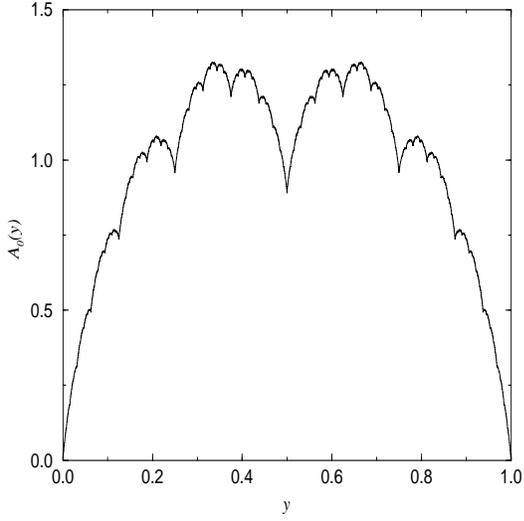,height=8cm,width=8cm}}
\caption{$A_0(y)$ computed with a cut-off value 
$k_{max} = 250$.}
\end{figure}

\begin{figure}[htb]
\centerline{\psfig{figure=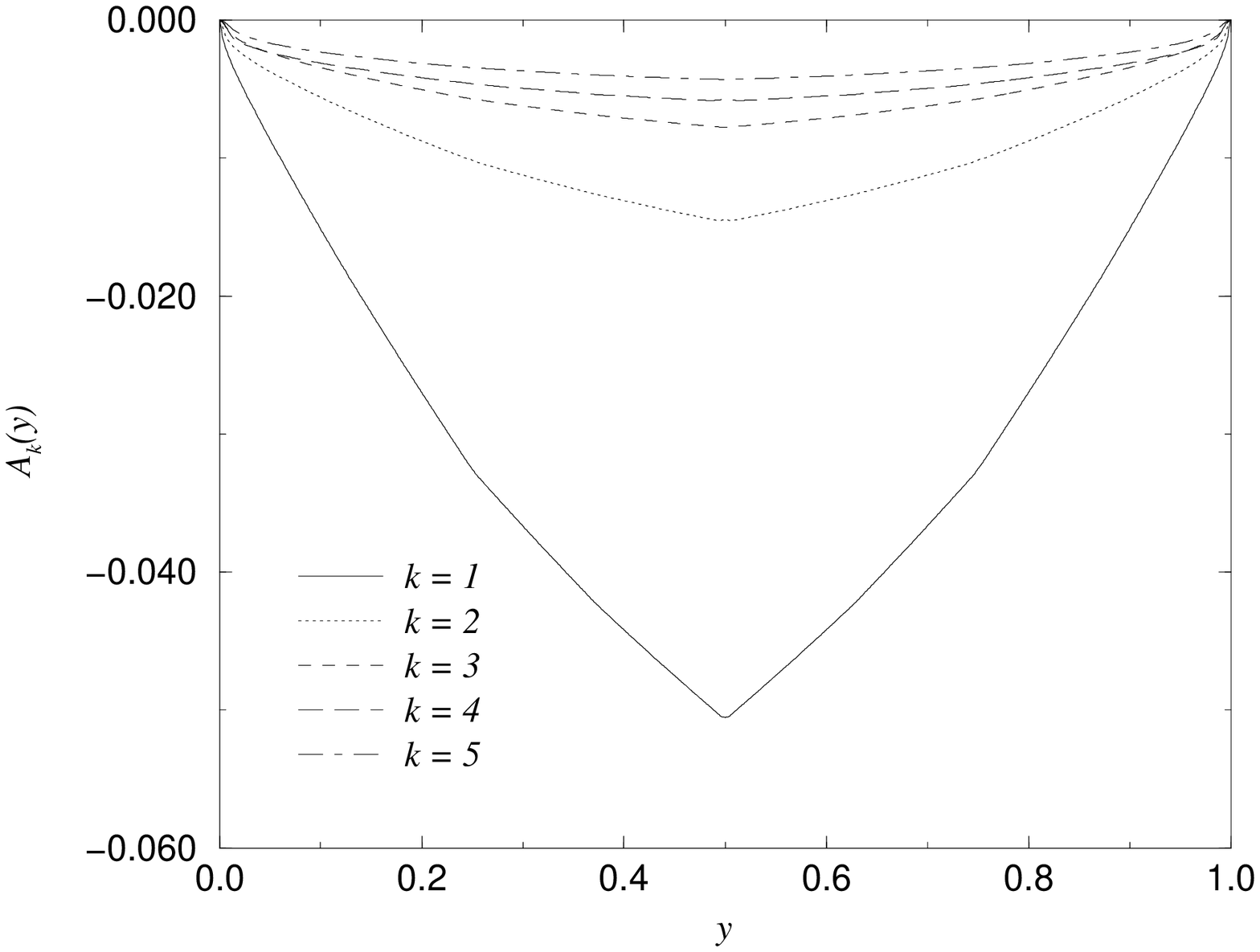,height=8cm,width=8cm}}
\caption{The first five $A_k(y)$ computed with a cut-off value 
$k_{max} = 250$.}
\end{figure}

\begin{figure}[htb]
\centerline{\psfig{figure=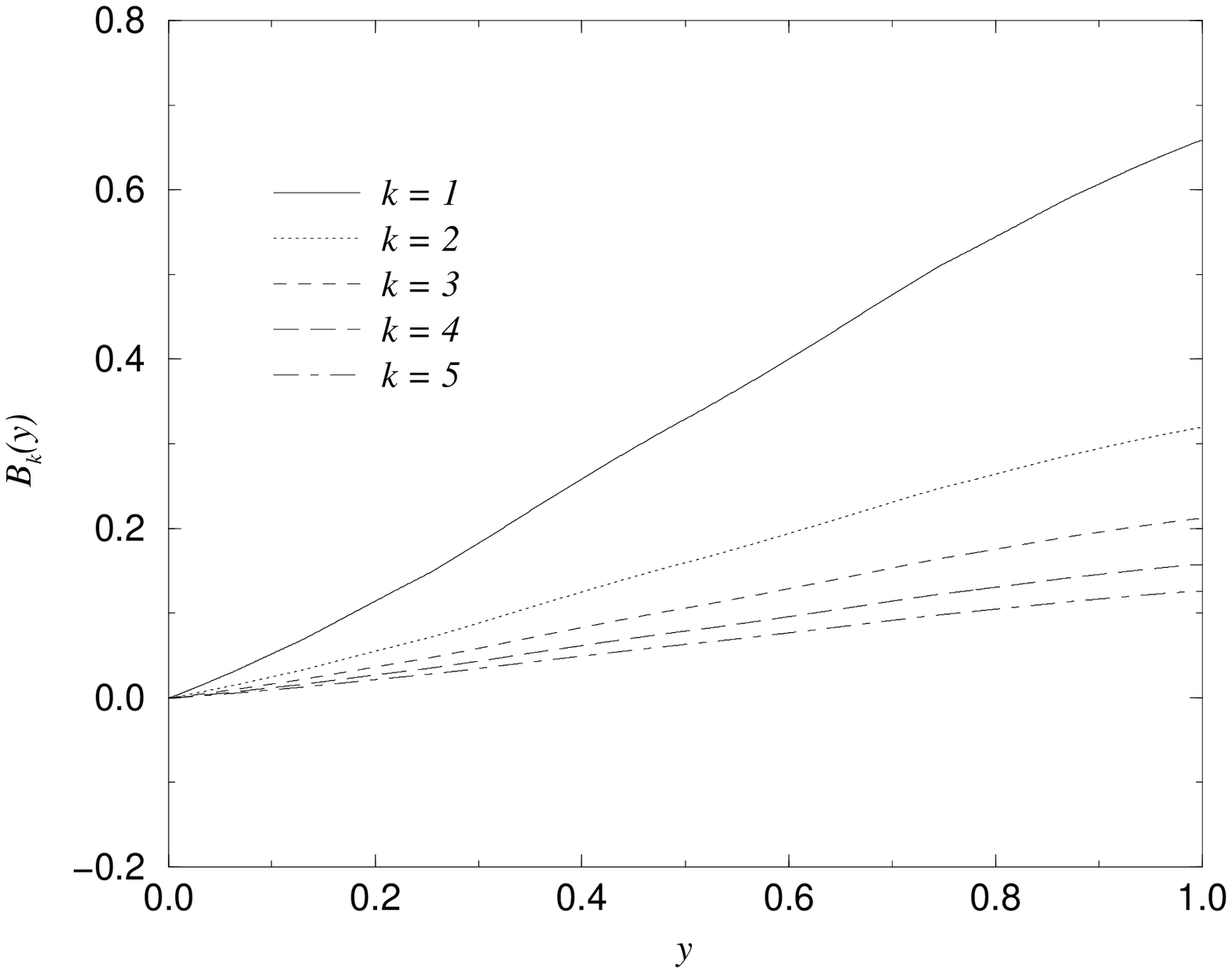,height=8cm,width=8cm}}
\caption{The first five $B_k(y)$ computed with a cut-off value 
$k_{max} = 250$.}
\end{figure}

\begin{figure}[htb]
\centerline{\psfig{figure=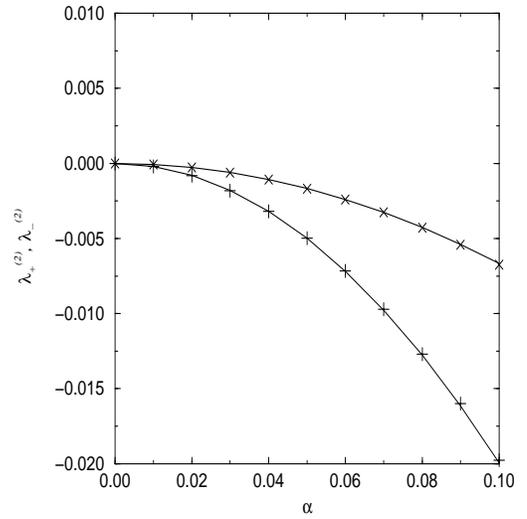,height=8cm,width=8cm}}
\caption{A comparison of the values of the second order corrections
of the Lyapuonv exponents computed from Eqs. (\ref{plyap}, \ref{nlyap}) 
(solid lines) and by following a trajectory of 10000 steps for 10 values of 
$\alpha$ ranging from 0 to .1 (x's correspond to the positive Lyapunov
exponent, +'s to the negative one).}
\end{figure}

\begin{figure}[htb]
\centerline{\psfig{figure=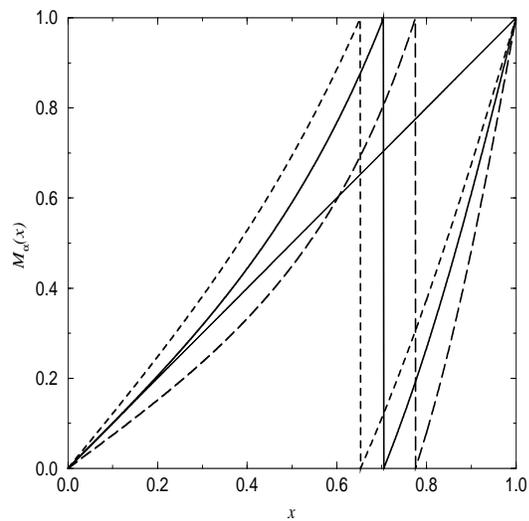,height=8cm,width=8cm}}
\caption{$M_\alpha$ projected along the $x$-interval for $\alpha = .5$ 
(dashed line),$\ln(2)$ (solid line),$1$ (long-dashed line). The origin goes 
from repelling to attractive as $\alpha$ increases above $\ln(2).$}
\end{figure}

\end{document}